\documentclass[twocolumn,showpacs,preprintnumbers,amsmath,amssymb]{revtex4}

\usepackage{graphicx}
\usepackage{dcolumn}
\usepackage{bm}

\begin{document}
\preprint{REV\TeX~4}

\title{Effect of microstructures on the electron-phonon interaction in the disordered metals Pd$_{60}$Ag$_{40}$}
\author{Y. L. Zhong}
\address{Department of Physics, National Tsing Hua University, Hsinchu 300, Taiwan}
\author{J. J. Lin}
\author{L. Y. Kao}
\address{Institute of Physics, National Chiao Tung University, Hsinchu 300, Taiwan}
\email{jjlin@cc.nctu.edu.tw}
\date{\today}

\begin{abstract}

Using the weak-localization method, we have measured the electron-phonon scattering times
$\tau_{ep}$ in Pd$_{60}$Ag$_{40}$ thick films prepared by DC- and RF-sputtering deposition
techniques. In both series of samples, we find an anomalous $1/\tau_{ep} \propto T^2\ell$
temperature and disorder dependence, where $\ell$ is the electron elastic mean free path. This
anomalous behavior cannot be explained in terms of the current concepts for the electron-phonon
interaction in impure conductors. Our result also reveals that the strength of the electron-phonon
coupling is much stronger in the DC than RF sputtered films, suggesting that the electron-phonon
interaction not only is sensitive to the total level of disorder but also is sensitive to the
microscopic quality of the disorder.

\end{abstract}
\maketitle

\section{Introduction}

The electron-phonon ($e$-ph) scattering time, $\tau_{ep}$, is one of the most important physical
quantities in metals and superconductors. For instance, it determines the dephasing (also called
the phase-breaking or decoherence) time for the electron wave function, the cooling time for an
electron gas, and the relaxation time for the order parameter in a superconductor. The $e$-ph
scattering time also plays a crucial role in the development of novel mesoscopic devices such as
sensitive low-temperature bolometers \cite{Gershen01}. The $e$-ph scattering time in the presence
of multiple (elastic) impurity scattering has been intensively calculated by several authors
\cite{Schmid86,Reizer86,Belitz87}, but the current understanding of the temperature and electron
elastic mean free path, $\ell$, dependences of $\tau_{ep}$ is still incomplete. In particular,
different temperature and disorder dependences of $\tau_{ep}$ have been reported, both
theoretically and experimentally \cite{Lin00,Lin02}. Recently, it was proposed that, in addition
to the dependence on the total level of disorder, the $T$ and $\ell$ dependence of $\tau_{ep}$
might be fairly sensitive to the microscopic quality of the disorder
\cite{Lin95,Lin98b,Sergeev00}. It has also been conjectured that the contribution due to the
Umklapp process of impurity scattering may be important \cite{GYWu01}.

In this work, we have fabricated two series of Pd$_{60}$Ag$_{40}$ thick films by DC-sputtering and
RF-sputtering deposition techniques. The palladium-silver alloys are chosen because Pd and Ag form
perfect fcc solid solution through the alloy series \cite{Laufer87}. Also, since the masses of the
Pd and Ag atoms are quite similar, the vibrational spectrum of the lattice does not change
significantly through the alloy series \cite{Dugdale}. The low-field magnetoresistances of our
films are measured at liquid-helium temperatures, and are compared with the weak-localization
theoretical predictions to extract the values of the $e$-ph scattering time. Our results for the
temperature and electron mean free path dependence of $\tau_{ep}$ and their implications are
described below.

\section{Experimental method}

Our films were prepared from a 99.995\% pure Pd$_{60}$Ag$_{40}$ (hereafter refereed to as PdAg)
target. Two series of thick films were fabricated, one by DC-sputtering and the other by
RF-sputtering deposition technique. The films were deposited onto glass substrates held at room
temperature. In both cases, a background pressure of $3 \times 10^{-6}$ torr was reached before an
argon atmosphere of $3.8 \times 10^{-3}$ torr was introduced to initiate the deposition process. A
same sputtering gun was used for these two deposition methods, but with the gun being connected to
either a DC or a RF power supply. The distance between the sputtering target and the glass
substrates was the same for both methods. The sputtering power was progressively adjusted to
``tune" the deposition rate, which resulted in different amounts of disorder, i.e., the residual
resistivities $\rho_0$ [= $\rho$(10\,K)], in the films. For the DC-sputtering (RF-sputtering)
case, the deposition rate was varied from 30 to 230 (19 to 333) $\rm\AA$/min, and values of
$\rho_{0}$ ranging from 281 to 183 (74 to 178) $\mu\Omega$ cm were obtained.

\begin{figure}
\includegraphics[angle=270,scale=0.32]{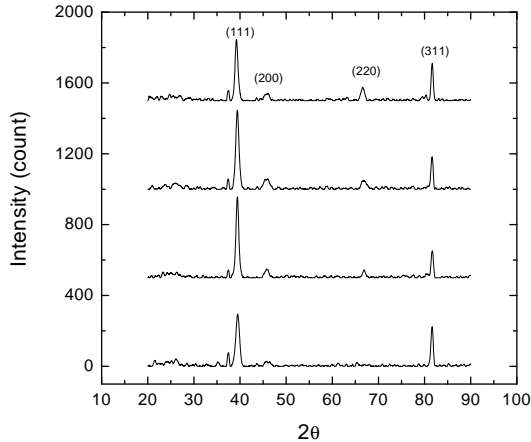}
\caption{\label{fig1} X-ray diffraction patterns for two DC (top two curves) and two RF (bottom
two curves) sputtered PdAg thick films. The sample resistivities $\rho$(300\,K) from top down are
189, 82, 79, and 132 $\mu \Omega$ cm. The diffraction intensity shows an arbitrary unit.}
\end{figure}

The sample structures of our films were carefully studied by performing the powder diffraction on
an MAC MXP18 x-ray diffractometer. The x-ray power was 10 kW and the scanning speed was 6 degrees
per minute. In all cases, we found our samples to reveal very similar diffraction patterns, which
clearly suggested that both the DC and RF sputtered films possessed the same fcc lattice structure
characteristic to that of the PdAg alloys. Representative x-ray diffraction patterns for two DC
and two RF sputtered films are shown in Fig. \ref{fig1}.

Most of our films had a thickness $t \gtrsim$ 4000 $\rm\AA$. This thickness ensured that the
weak-localization effects were three-dimensional in our samples. It also ensured that the thermal
phonons were unambiguously three-dimensional, i.e., the wavelength of the most probable thermal
phonons was always smaller than the film thickness at our measurement temperatures. This latter
condition greatly eliminated any complications that might result from phonon confinement effects.
(In reduced-dimensional systems, modifications to the phonon spectrum and the speed of sound might
be significant, which could lead to non-straightforward temperature and disorder behavior of
$\tau_{ep}$.)

Our values of the diffusion constant, $D$, were evaluated through the Einstein relation
$\rho_0^{-1} = D e^{2} N(0)/(1+ \lambda)$, where $N(0)$ is the electronic density of states at the
Fermi level, and $\lambda$ is the $e$-ph mass enhancement factor. The values of $N(0)$ were
calculated from the independently determined electronic specific heat: $\gamma T = \frac13 \pi^2
k_B^2 N(0)T$. For Pd$_{60}$Ag$_{40}$, $\lambda \simeq$ 0.43, and $\gamma \simeq$ 3.3 mJ/mol K$^2$
\cite{Laufer87}. Then, we obtained $D \approx (100/\rho_0)$ cm$^2$/s, where $\rho_0$ is in $\mu
\Omega$ cm. Table \ref{t1} lists the values of the relevant parameters for our films studied in
this work.

\begin{table*}
\caption{\label{t1} Values of the relevant parameters for Pd$_{60}$Ag$_{40}$ thick films. The
samples end with (without) an dagger denote films prepared by DC- (RF-)sputtering deposition
technique. $t$ is the film thickness. $\rho_0$ is the resistivity at 10 K. $D$ is the diffusion
coefficient. The values of $k_F\ell = 3mD/\hbar$ are computed by assuming the free electron mass
of $m$, where $k_F$ is the Fermi wave number. $\tau_\phi^0$ is the fitted electron dephasing time
as $T \rightarrow 0$. $A_{ep}$ and $p$ are the fitted strength of $e$-ph coupling and effective
exponent of temperature, respectively, in $1/\tau_{ep} = A_{ep} T^p$.}

\begin{ruledtabular}
\begin{tabular}{lccccccc}
Sample & $t$\,($\rm\AA$) & $\rho_0$\,($\mu\Omega$ cm) & $D$\,(cm$^2$/s) & $k_F\ell$ &
$\tau_\phi^0$\,($10^{-10}$ s) & $A_{ep}$\,($10^8$ s$^{-1}$ K$^{-p}$) & $p$ \\ \hline
PdAg11$^\dag$ & 3900 & 281 & 0.36 & 0.93 & 2.8 & 1.5 & 2.3$\pm$0.1 \\
PdAg15$^\dag$ & 5100 & 235 & 0.43 & 1.1 & 1.1 & 2.4 & 1.9$\pm$0.2 \\
PdAg08$^\dag$ & 3900 & 224 & 0.45 & 1.2 & 6.7 & 2.0 & 2.4$\pm$0.1 \\
PdAg12$^\dag$ & 4800 & 183 & 0.55 & 1.4 & 3.7 & 2.8 & 2.2$\pm$0.1 \\
PdAg18 & 5000 & 178 & 0.56 & 1.5 & 4.3 & 0.40 & 2.4$\pm$0.1 \\
PdAg17 & 4500 & 101 & 0.99 & 2.6 & 1.6 & 2.2 & 2.2$\pm$0.1 \\
PdAg19 & 4000 & 98 & 1.0 & 2.6 & 1.6 & 2.9 & 2.3$\pm$0.1 \\
PdAg14 & 4100 & 90 & 1.1 & 2.9 & 1.1 & 2.5 & 2.2$\pm$0.2 \\
PdAg16 & 3300 & 74 & 1.3 & 3.5 & 0.97 & 3.6 & 2.2$\pm$0.1
\end{tabular}
\end{ruledtabular}
\end{table*}

\section{Results and Discussion}

\begin{figure}
\includegraphics[angle=270,scale=0.32]{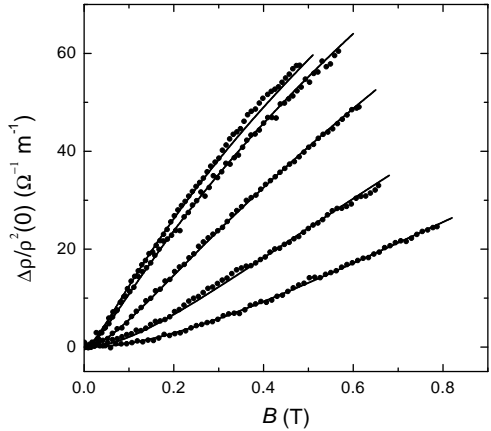}
\caption{\label{fig2} Normalized magnetoresistivities as a function of magnetic field for the
PdAg17 thick film at (from top down) 1.0, 3.0, 6.0, 9.0, and 14.0 K. The solid curves are the
three-dimensional weak-localization theoretical predictions.}
\end{figure}

The normalized magnetoresistivities, $\triangle \rho (B)/\rho^2(0) = [\rho (B)-\rho(0)]/\rho^2
(0)$, for the PdAg17 thick film at several temperatures are plotted in Fig. \ref{fig2}. The
symbols are the experimental data and the solid curves are the three-dimensional weak-localization
theoretical predictions \cite{Fuku81}. It is clearly seen that the weak-localization predictions
can describe well our experimental data. Therefore, the electron dephasing time $\tau_\phi$, which
is the key parameter in the weak-localization theory, can be reliably extracted. Since PdAg has a
very strong spin-orbit scattering, $\tau_\phi$ is the {\em only} adjusting parameter in the
comparison of the theory with experiment. (That the spin-orbit scattering is strong in PdAg is
evident in the shape of the positive magnetoresistivity curves shown in Fig. \ref{fig2}.) The
details of our data analysis procedure was discussed previously \cite{Lin94}.

In {\em three} dimensions, the total electron dephasing rate that governs the weak-localization
effects is given by \cite{Fuku81}
\begin{equation}
\frac{1}{\tau_\phi (T,\ell)} = \frac{1}{\tau_\phi^0 (\ell)} + \frac{1}{\tau_{ep}(T,\ell)} \,,
\label{eq1}
\end{equation}
where $\tau_\phi^0 = \tau_\phi (T \rightarrow 0)$ depends very weakly on temperature, if at all,
and is called the saturated dephasing time. Whether $\tau_\phi^0$ should reach a finite or an
infinite value as $T \rightarrow 0$ is currently under vigorous experimental and theoretical
investigations \cite{Lin02}. At finite temperatures, the dominating inelastic electron process in
three dimensions is solely due to the $e$-ph scattering, while the Nyquist electron-electron
scattering is negligibly small \cite{Lin02,Lin98,Gershen99}. Usually, one writes $1/\tau_{ep} =
A_{ep}T^p$ over the limited temperature range accessible in a typical experiment, where $A_{ep}$
characterizes the strength of the $e$-ph coupling, and $p$ is an effective exponent of
temperature. According to current understanding, $p$ lies between 2 and 4
\cite{Schmid86,Reizer86,Belitz87,Sergeev00}.

\begin{figure}
\includegraphics[scale=0.32]{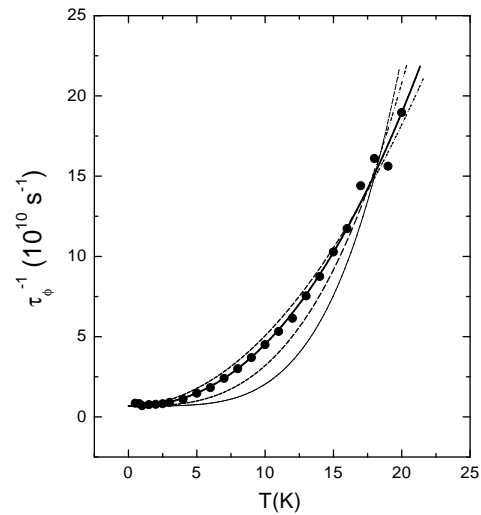}
\caption{\label{fig3} Electron dephasing rate as a function of temperature for the PdAg17 thick
film. The thick solid curve drawn through the data points is a least-squares fit to Eq.
(\ref{eq1}), using $p$ as a free parameter. The dotted, dashed, and thin solid curves are
least-squares fits to Eq. (\ref{eq1}) with $p$ fixed at 2, 3, and 4, respectively (see text).}
\end{figure}

The extracted $\tau_\phi (T)$ between 0.5 and 20 K for each of our films is least-squares fitted
to Eq. (\ref{eq1}), and the fitted values of the relevant parameters ($\tau_\phi^0$, $A_{ep}$, and
$p$) are listed in Table \ref{t1}. Figure \ref{fig3} shows a plot of the variation of
$1/\tau_\phi$ with temperature for the PdAg17 thick film. The symbols are the experimental data.
The thick solid curve drawn through the data points is obtained with $\tau_\phi^0$, $A_{ep}$, and
$p$ as free parameters. In this case, we obtain a temperature exponent $p = 2.2 \pm 0.1$. For
comparison, we have also least-squares fitted the measured $1/\tau_\phi$ with Eq. (\ref{eq1}), but
with $p$ fixed at an integer value of either 2, 3, or 4 (while allowing $\tau_\phi^0$ and $A_{ep}$
to vary). The dotted, dashed, and thin solid curves in Fig. \ref{fig3} plot the fitted results
with $p$ = 2, 3, and 4, respectively. It is clearly seen that our temperature dependence of
$1/\tau_{ep}$ can be best described with an exponent $p$ equal or close to 2. In fact, we have
found that the temperature behavior of $\tau_{ep}$ for all films listed in Table \ref{t1} is very
similar, i.e., $1/\tau_{ep}$ demonstrates an {\em essentially quadratic} temperature dependence.

\begin{figure}
\includegraphics[angle=270,scale=0.32]{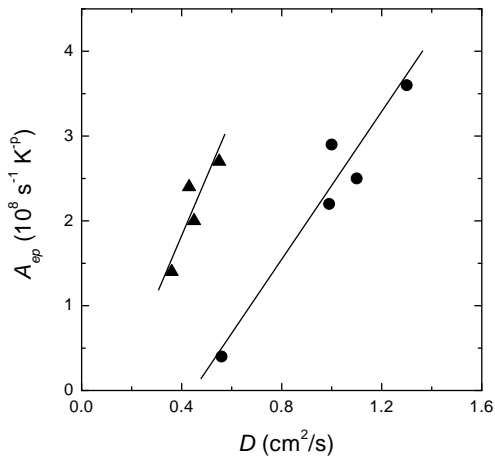}
\caption{\label{fig4} The strength of $e$-ph coupling $A_{ep}$ as a function of diffusion constant
for DC (triangles) and RF (circles) sputtered PdAg thick films. The straight lines drawn through
the data points are guides to the eye.}
\end{figure}

Inspection of Table \ref{t1} indicates that, for either DC or RF sputtered films, the value of
$A_{ep}$ decreases with increasing level of disorder ($\rho_0$) in the sample. Figure \ref{fig4}
plots the fitted $A_{ep}$ as a function of the diffusion constant. Clearly, ones sees that
$A_{ep}$ varies {\em linearly} with $D$, implying that $1/\tau_{ep} \propto D \propto \ell$. It
should be noted that, if we plot $1/\tau_{ep}$ as a function of the measured $\rho_0^{-1}$, we
also observe a linear variation, i.e., $1/\tau_{ep} \propto \rho_0^{-1} \propto \ell$. Such a
linearity of $A_{ep}$ with $\ell$ holds for {\em both} series of films. Quantitatively, however,
the values of $A_{ep}$ (for a given disorder) for the DC and RF sputtered films are very
different. For example, while the DC sputtered PdAg12 and the RF sputtered PdAg18 thick films have
essentially the same $\rho_0$, their values of $A_{ep}$ differ by several times. Moreover, Fig.
\ref{fig4} reveals that the slope of the linearity is about a factor of 2 larger in the DC than RF
sputtered films. Since the x-ray diffraction studies demonstrate that the crystal structures are
quite similar for both series of films (Fig. \ref{fig1}), the differences in the values of
$A_{ep}$ and the variation of $A_{ep}$ with $\ell$ strongly imply that the $e$-ph interaction must
be very sensitive to the microstructures of the samples. The subtle difference in the
microstructures in these two series of films may result from the different ways of sample
preparation.

Taken together, Fig. \ref{fig3} and Fig. \ref{fig4} demonstrate that the $e$-ph scattering in PdAg
possesses an {\em anomalous} temperature and disorder dependence of $1/\tau_{ep} \propto T^2
\ell$. This dependence is insensitive to the fabrication method. Such a $T^2 \ell$ behavior is
totally unexpected, even qualitatively, in terms of the current theoretical concepts for the
$e$-ph interaction in impure conductors. According to the ``orthodox" $e$-ph interaction theory
for disordered metals \cite{Schmid86,Reizer86,Belitz87}, that assumes a coherent motion of the
impurity atoms with the deformed lattice atoms at low temperatures, one should expect a $T^4 \ell$
dependence. Recently, it was speculated that, in real metals containing heavy (light) impurities
and tough boundaries, the impurity and/or boundary atoms might not move in phase with the lattice
atoms \cite{Lin95}. The first calculations in consideration of this effect have been done by
Sergeev and Mitin \cite{Sergeev00}. They found that even a small amount of ``static" potential
scatterers drastically changes the $e$-ph-impurity interference, and the relaxation rate is
proportional to $T^2 {\cal L}^{-1}$, where ${\cal L}$ is the electron mean free path with respect
to the static impurities (${\cal L} \gg \ell$). Experimentally, a $T^4$ temperature dependence has
been observed very recently in disordered Hf and Ti thin films \cite{Gershen01}. (A $T^4$
dependence had been previously observed in Bi thin films over a very limited temperature range of
0.6$-$1.2 K \cite{Komnik94}.) However, to the best of the authors' knowledge, the combined
$T^4\ell$ law has never been confirmed in real conductors thus far. On the other hand, a distinct
$T^2 \ell^{-1}$ dependence has been observed in Ti$_{1-x}$Al$_x$ \cite{Lin95} and Ti$_{1-x}$Sn$_x$
alloys \cite{Lin98b}. Previously, a $T^2 \ell$ dependence was independently found in AuPd {\em
thick} films ($t \gtrsim$ 4000 $\rm\AA$) \cite{Lin98}, and Nb {\em thin} films ($t \lesssim$ 200
$\rm\AA$) \cite{Gershen90}. In the present case of PdAg thick films, the masses of the Pd and Ag
atoms are quite similar, and the films are three-dimensional. Therefore, it is not clear how the
Sergeev-Mitin theory evoking heavy (light) impurities and tough boundaries can apply to this case.

The criterion for the $e$-ph interaction to satisfy the dirty-limit condition is $q_T \ell \ll 1$,
where the wave number of the thermal phonons $q_T \approx k_BT /\hbar v_s$, and $v_s$ is the speed
of sound. Taking $v_s \approx$ 2600 m/s \cite{vs} and $\ell \approx$ 2$-$8 $\rm\AA$, we obtain
$q_T T \approx$ (0.01$-$0.04)$T$ for our PdAg thick films. The phase-breaking lengths $\sqrt{D
\tau_\phi}$ in our films are calculated to be 690$-$1500 $\rm\AA$ at 2 K. (The dephasing length
essentially saturates below about 2 K.) This length scale justifies the use of three-dimensional
weak-localization theory to describe our experimental magnetoresistivities.

\section{Conclusion}

We have measured the $e$-ph scattering time $\tau_{ep}$ in DC and RF sputtered PdAg thick films.
In both series of films, we observe an anomalous $1/\tau_{ep} \propto T^2\ell$ temperature and
disorder dependence. Moreover, the $e$-ph coupling is found to be much stronger in the DC than RF
sputtered films. This observation strongly indicates that the $e$-ph interaction not only is
sensitive to the total level of disorder but also is sensitive to the microscopic quality of the
disorder. These results pose a new theoretical challenge.

\begin{acknowledgments}
We are grateful to V. Mitin, A. Sergeev, and G. Y. Wu for valuable discussions. This work was
supported by the Taiwan National Science Council through Grant Nos. NSC90-2112-M-009-037 and
NSC90-2119-M-007-004.
\end{acknowledgments}


\begin{references}

\bibitem{Gershen01} M.E. Gershenson, D. Gong, T. Sato, B.S. Karasik, and A.V. Sergeev, Appl. Phys.
Lett. {\bf 79}, 2049 (2001).

\bibitem{Schmid86} J. Rammer and A. Schmid, Phys. Rev. B {\bf 34}, 1352 (1986).

\bibitem{Reizer86} M.Yu Reizer and A.V. Sergeyev, Zh. Eksp. Teor. Fiz. {\bf 90}, 1056 (1986)
[Sov. Phys. JETP {\bf 63}, 616 (1986)].

\bibitem{Belitz87} D. Belitz, Phys. Rev. B {\bf 36}, 2513 (1987).

\bibitem{Lin00} J.J. Lin, Physica B {\bf 279}, 191 (2000).

\bibitem{Lin02} See, for a recent review, J.J. Lin and J.P. Bird, J. Phys.: Condens. Matter
{\bf 14}, R501 (2002).

\bibitem{Lin95} J.J. Lin and C.Y. Wu, Europhys. Lett. {\bf 29}, 141 (1995).

\bibitem{Lin98b} C.Y. Wu, W.B. Jian, and J.J. Lin, Phys. Rev. B {\bf 57}, 11232 (1998).

\bibitem{Sergeev00} A. Sergeev and V. Mitin, Phys. Rev. B {\bf 61}, 6041 (2000); Europhys. Lett.
{\bf 51}, 641 (2000).

\bibitem{GYWu01} W. Jan, G.Y. Wu, and H.S. Wei, Phys. Rev. B {\bf 64}, 165101 (2001); W. Jan and
G.Y. Wu, J. Phys.: Condens. Matter {\bf 13}, 10925 (2001).

\bibitem{Laufer87} P.M. Laufer and D. A. Papaconstantopoulos, Phys. Rev. B {\bf 35}, 9019 (1987).

\bibitem{Dugdale} J.S. Dugdale, {\it The Electrical Properties of Metals and Alloys}
(Edward Arnold, London, 1977).

\bibitem{Fuku81} H. Fukuyama and K. Hoshino, J. Phys. Soc. Jpn. {\bf 50}, 2131 (1981).

\bibitem{Lin94}  C.Y. Wu and J.J. Lin, Phys. Rev. B {\bf 50}, 385 (1994).

\bibitem{Lin98} Y.L. Zhong and J.J. Lin, Phys. Rev. Lett. {\bf 80}, 588 (1998).

\bibitem{Gershen99} M.E. Gershenson, Ann. Phys.-Leipzig {\bf 8}, 559 (1999).

\bibitem{Komnik94} Yu.F. Komnik, V.Yu. Kashirin, B.I. Belevtsev, and E.Yu. Beliaev, Phys. Rev. B
{\bf 50}, 15298 (1994).

\bibitem{Gershen90} E.M. Gershenzon, M.E. Gershenzon, G.N. Gol'tsman, A.M. Lyul'kin, A.D.
Semenov, and A.V. Sergeev, Zh. Eksp. Teor. Fiz. {\bf 97}, 901 (1990) [Sov. Phys. JEPT {\bf 70},
505 (1990)].

\bibitem{vs} {\it The Practicing Scientist's Handbook}, edited by A. J. Moses (van
Nostrand/Reinhold, New York, 1978); W.C. McGinnis and P.M. Chaikin, Phys. Rev. B {\bf 32}, 6319
(1985).

\end{references}

\end{document}